\documentclass{sig-alternate-mama}
\usepackage{floatflt,graphicx}
\usepackage{subfigure}

\begin{document}

%

\title{Cost Sharing in Social Community Networks}

\numberofauthors{2}

\author{
%
\alignauthor Ranjan Pal\\
                              \affaddr{Department of Computer Science}\\
                               \affaddr{University of Southern California}\\
                               \email{rpal@usc.edu}
\and
\alignauthor Aravind Kailas\\
                               \affaddr{Department of Electrical Engineering}\\
                               \affaddr{University of North Carolina, Charlotte}\\
                               \email{Aravind.Kailas@uncc.edu}}

\maketitle
\begin{abstract}
Wireless social community networks (WSCNs) is an emerging technology that operate in the unlicensed spectrum and have been created as an alternative to cellular wireless networks for providing low-cost, high speed wireless data access in urban areas. WSCNs is an upcoming idea that is starting to gain attention amongst the civilian Internet users. By using \emph{special} WiFi routers that are provided by a social community network provider (SCNP), users can effectively share their connection with the neighborhood in return for some monthly monetary benefits. However, deployment maps of existing WSCNs reflect their slow progress in capturing the WiFi router market. In this paper, we look at a router design and cost sharing problem in WSCNs to improve deployment. We devise a \emph{simple to implement}, \emph{successful}\footnote{a mechanism is successful if it achieves its intended purpose. For example in this work, a successful mechanism would help install routers in a locality}, \emph{budget-balanced}, \emph{ex-post efficient}, and \emph{individually rational}\footnote{a mechanism is individually rational if the benefit each agent obtains is greater than its cost.} auction-based mechanism that generates the \emph{optimal} number of features a router should have and allocates costs to residential users in \emph{proportion} to the feature benefits they receive. Our problem is important to a new-entrant SCNP when it wants to design its multi-feature routers with the goal to popularize them and increase their deployment in a residential locality. Our proposed mechanism accounts for heterogeneous user preferences towards different router features and comes up with the optimal \emph{(feature-set, user costs)} router blueprint that satisfies each user in a locality, in turn motivating them to buy routers and thereby improve deployment.

\emph{Keywords:} wireless social community networks, multi-feature routers, auction mechanism, cost-sharing
\end{abstract}
\section{Introduction}
The last few years has seen the rapid increase in demand for high-speed wireless data services. These services that are traditionally offered by cellular service providers operate in the licensed band of the radio spectrum. The cellular service providers guarantee high quality of service (QoS) and good coverage but charge high prices to users due to their substantial investments, both for deploying and maintaining the network infrastructure, and for licensing the wireless spectrum.

A low-cost, high-speed alternative called \emph{wireless social community networks} (WSCNs) has emerged in the last five years for providing cheap and ubiquitous WiFi access to users. These networks operate in unlicensed spectrum bands and use WiFi access points that are operated and maintained by community members to provide data service. A well-known example of a company that provides WSCN services is FON \emph{(http://www.fon.com)}, which is a worldwide WiFi community network funded by commercial organizations like Google, Skype, and Free \emph{(http://www.free.fr)}. Users in a WSCN buy special multi-feature\footnote{Routers sold by FON act like a virtual PC. They can upload video on YouTube, load photos on Picasa and Facebook, can download Internet files and P2P data, etc., even when the PC is off, have some special hardware components, etc. Presently, these routers are priced in the range of 40 to 80 dollars per unit and can be purchased online from sites like Amazon.} WiFi routers from a social network community provider (SCNP) like FON and share their bandwidth \emph{securely} with other users around their locality in return for some monthly monetary benefits. FON members \emph{(Foneros)} can freely access WiFi anywhere in the world near a locality where FON members are present, whereas non-FON members need to pay the SCNP for using the bandwidth of a FON member. Thus, a WSCN provides a platform for mobile and ubiquitous wireless access.

Although the promise shown by the wireless social networks is good, current deployment maps indicate that WSCNs are yet to capture the WiFi router market, i.e., the router deployment rate is quite slow. There could be multiple reasons for such a trend: 1) users may not be well informed about the technology, 2) they may be skeptical about the monetary benefits earned in comparison to the services they provide (bandwidth sharing), 3) there could be certain ISP policies that could prevent the SCNP's from widely marketing their routers\footnote{Mobile Internet users may want to access WSCN bandwidth and pay the SCNP instead of using a traditional Internet service provider (ISP) connection. This may not be acceptable to ISP's for commercial profit reasons. Thus, the SCNP's and ISP's may settle a contract that lays down certain policies of operation.}, and 4) common Internet users may just be happy with their traditional ISP connection and may not be bothered about sharing their WiFi with others. All the above mentioned factors contribute to low router deployment and in turn reduce ubiquitous network coverage and quality of service.

In this paper, we address a router design and cost sharing problem in WSCNs. Our problem is important to a \emph{new-entrant} SCNP when it wants to design its multi-feature\footnote{The features may be, both hardware and software in nature.} routers with the goal to popularize them and increase their deployment in a residential locality. Since the users in a residential locality are heterogeneous in nature w.r.t feature preferences, a feature may not be equally viable to two users in a network. Thus, without any proper incentive mechanism, it is quite likely that many features will not seem important enough for users buy, and as a result the SCNP may end up with a low deployment of their routers. We devise a \emph{simple to implement}, \emph{successful}, \emph{budget-balanced}, \emph{ex-post efficient}, and \emph{individually rational} auction-based mechanism that generates the \emph{optimal} number of features a router should have, and allocates costs to residential users in \emph{proportion} to the feature benefits they receive. Our proposed mechanism accounts for heterogeneous user preferences towards different router features and comes up with the optimal \emph{(feature-set, user costs)} router blueprint that satisfies each user in a locality, in turn motivating them to buy routers and thereby improve deployment. To the best of our knowledge, this is the first work of its kind in the area of wireless social community networks.

\emph{Related Work:} Research in wireless social community networks is relatively new. Efstathiou et.al propose a charging model for wireless community networks using the concept of user reciprocation \cite{efp}. In a recent series of works \cite{mffmh} \cite{mmh}, Hubaux et.al., have devised static and dynamic pricing strategies in WSCNs for reaching high network coverage. They have also studied the evolution and market share of WSCNs using non-cooperative game theory. The authors observe that 1) the dynamics of the community depends on initial coverage, the subscription fee, user preferences for coverage, and the WiFi access points density, and 2) for the game where a mobile user can choose between the services of a licensed band operator and a SCNP, there exists a Nash equilibrium for specific distribution of user preferences. The Nash equilibrium characterizes the number of users that should subscribe to each type of service. Although the authors make a good study of a WSCN model, their strategies are difficult to implement in practice, due to subscription fee changes in different time slots. Variable fees imply the possibility of user switching, and this might not be a reality at present. The authors also do not consider the more basic problem of efficient router deployment and user buying motivation, which in itself is the first step to achieving high coverage.

\emph{Our Research Contribution}
\begin{itemize}
\item  We devise a \emph{simple to implement}, \emph{successful}, \emph{budget-balanced}, \emph{ex-post efficient}, and \emph{individually rational} auction-based mechanism that generates the \emph{optimal} number of features a WSCN router should have, and allocates costs to residential users in \emph{proportion} to the feature benefits they receive. Our proposed mechanism accounts for heterogeneous user preferences towards different router features and comes up with the optimal \emph{(feature-set, user costs)} router blueprint that satisfies each user in a locality, in turn motivating them to buy routers and thereby improve deployment.
\end{itemize}

The rest of the paper is organized as follows. In Section 2, we define our problem setup. Here we state our problem assumptions and explain mathematical notations to be used for modeling and analysis. In Section 3, we define certain mathematical properties of cost sharing rules used in our auction mechanism. We propose our auction mechanism in Section 4. We conclude our paper in Section 5.

\section{Problem Setup}
In this section, we state our problem assumptions and explain the mathematical notations used for modeling and analysis.
\subsection{Assumptions}
A wireless social network provider (SCNP) wants to market its business in a residential locality\footnote{A residential locality could be a residential complex like a condominium or apartment complex of 100's of apartments owned by an organization. This type of residential setup is very common in India and south-east asian countries. The residents of a condominium share common features and may be on interacting terms with one another.} by designing and selling routers that incorporate certain number of hardware and software features which could form the core of next generation wireless router design. We assume here that wireless social community networks is a new technology, and that the SCNPs are in the process of entering the market by designing their first set of multi-feature routers to meet the needs of the locality\footnote{Each locality may result in routers with different feature sets. However, at this juncture we assume for simplicity that people are generally uniformly distributed with respect to their needs, and as a result every locality will have the same needs on average.}. Each feature is advertised for (i.e., through newspapers, pamphlets, etc.,) and analyzed on a cost-benefit basis before being put into manufacture. Each user in a given geographical locality may or may not be interested in buying such routers. However, those who are interested are quite keen on getting proper information about the new technology at hand and its benefits. Given that most civilian users are naive about upcoming technology, they prefer to exchange information amongst the peers in the locality by mere personal interaction or using \emph{Facebook} like social websites and gauge the societal benefit as well as their own benefits. Typical information exchange would include information such as 1) user costs (buying hardware and  setup costs), 2) perceived user and societal benefits, 3) market information for newly advertised product, 4) hardware (technological) information, 5) network topologies surrounding the locality users, and 6) Internet service provider (ISP) policies. In this paper, we do not deal with the mechanism to exchange messages through Facebook. We believe that with the development of systems networking applications over time, efficient and secure message passing in social networks would not be an issue. Finally, we assume that the users in a locality who are interested in buying network routers are keen on societal welfare as well as their own individual welfare.
\subsection{Mathematical Notations}
In this section, we define and explain the basic and major mathematical notations that we use throughout the paper. Specific notations pertaining to a particular section will be discussed within the section itself.

$n$ - number of users in a geographical locality that are interested in the new multi-feature router product advertised by an SCNP. \\

$jb_{i}(k)$ - the estimate of the \emph{joint benefit} perceived by user $i$ of incorporating a particular feature $k$ in the router. Each user has its \emph{own} estimate that he obtains by communicating with its locality peers on Facebook regarding information specified in Section 2.1. We assume that the estimate of the joint benefit is a \emph{non-increasing} function in the number of features, i.e ., the marginal joint benefit monotonically decreases with the increase in the number of features. As an example, the estimated joint benefit to manufacture feature 1 is greater than or equal to the estimated joint benefit to manufacture feature 2. This assumption is realistic as router features generally have a \emph{sorted order of preferences} in a community w.r.t joint benefits. The estimates for each user are known to a third party called a `social planner' who is a mediator between the SCNP and the residential users. The estimate of the joint benefit can be considered as a measure of the \emph{overall social utility} of a particular router feature as perceived by a particular user $i$. \\

$ib_{i}(k)$ - user $i's$ \emph{reported} value of the valuation of a router feature $k$. This is a user's individual valuation as reported to the social planner. It may not be equal to a user's actual perceived utility for a feature. As in the case of estimated joint benefits, it is also a non-increasing function in the number of features. The reported valuation can be considered as a measure of the \emph{reported individual utility} of a particular router feature as decided by a particular user $i$. \\

$\overline{ib}(k)$ - maximum reported valuation of router feature $k$. \\

$\overline{jb}(k)$ - perceived valuation of the user who wins the bid for feature $k$. \\

$pb_{i}(k)$ - user $i's$ \emph{actual} value of the valuation of a router feature $k$. This is a user's true individual valuation, which he \emph{does not} report to the social planner. As in the case of estimated joint benefits, it is also a non-increasing function in the number of features. The reported valuation can be considered as a measure of the \emph{actual individual utility} of a particular router feature as perceived by a particular user $i$. We emphasize that once feature $k$ is manufactured, $pb_{i}(k)$ is the individual utility to user $i$. \\

$c(k)$ - the cost to an SCNP to manufacture router feature $k$. We assume the cost to be non-decreasing in the number of features. i.e ., the marginal manufacture cost monotonically increases with the increase in the number of features. As an example, the estimated cost to manufacture feature 1 is greater than or equal to the estimated cost to manufacture feature 2. This implies that given a sorted order of feature preferences w.r.t the whole locality, it takes an SCNP more cost to manufacture low preference features. The assumption is realistic because the cost may not just be manufacturing costs to build feature $k$ but could also include costs in the form of time, prospective profits, market impact, etc., which might reduce the level of importance of SCNPs to manufacture low preference features. \\

$\pi$ - a family of cost sharing rules. It is a per-feature extension of the family proposed in \cite{jm}, which considers only one feature. We explain more about cost sharing rules in Sections 3 and 4. \\

$\Phi^{PF}$ - the set of all per-feature extensions, $pi$. \\

$\varphi_{k}$ - a per-feature proportional cost sharing rule for feature $k$. It is a per-feature extension of the rule proposed in \cite{jm} for a single feature.\\

For modeling simplicity, we assume $jb_{i}(k)$, $ib_{i}(k)$, $pb_{i}(k)$, and $c(k)$ to have the same measurable units. This assumption need not be necessarily true, and it is an important open modeling question in cyber-economic systems to model these parameters with the most appropriate units.

\section{Properties of Cost Sharing Rules}
In this section, we define the mathematical properties that our cost sharing rules possess. The properties define the nature of our cost sharing rules in light of users sharing benefits and providers (SCNPs) expending cost to manufacture features. We modify the properties of the family, $\Phi$, of cost sharing rules that were stated in \cite{jm} to account for a router with multiple-features\footnote{the authors in \cite{jm} just consider one feature.}. We emphasize that a feature $k$ is manufactured iff $\sum_{i = 1}^{n}pb_{i}(k) \geq c(k)$, i.e., there is no use for a WSCN user to buy a router with feature $k$ if the user's actual individual utility is lesser than the cost to manufacture feature $k$. Let $F$ be the largest $k$ such that $\sum_{i = 1}^{n}pb_{i}(k) > c(k)$. Then exactly $F$ features will be manufactured. Let $pb_{i} = (pb_{i}(1),.....,pb_{i}(F))$ and $C = (c(1),......,c(F))$. We define our cost-sharing rule over the domain $D^{F}$, where $D^{F} = \{(pb_{1},.....,pb_{n}): ib_{i} \geq 0; \sum_{i = 1}^{n}pb_{i}(k) > c(k), \forall k \leq F\}$.

A \emph{per-feature} extension of $\Phi$ is a mapping $\pi$ defined over $D^{F}$ and with range [0, $\sum_{k = 1}^{F} c(k)$], satisfying the following
\[\pi(pb_{1},.....,pb_{n}; C) = \sum_{k = 1}^{F} \varphi_{k}(pb_{1}(k),.....,pb_{n}(k); c(k)); \varphi_{k}\,\epsilon\,\pi\, \forall k;\]
where $\varphi_{k}$ is a cost-sharing rule in the per-feature extension family of cost sharing rules devised by Moulin et.al. in \cite{jm}\footnote{any cost sharing rule in the family of rules in \cite{jm} follows \emph{monotonicity} properties mentioned in \cite{jm}. In this section, we extend the monotonicity properties mentioned in \cite{jm} to account for multiple features.}. The per-feature monotonicity property extensions from those in \cite{jm} are as follows.
\[\varphi_{k}(pb_{i}(k);pb_{-i}(k); c(k)) \mbox{ is non-increasing in } pb_{-i}(k)\]
and $\forall \lambda > 0$ we have
\[\varphi_{k}(pb_{i}(k) - \lambda; pb_{j \ne i} + \lambda,...,pb_{n}(k)) \ge \varphi_{k}(pb_{i}(k);pb_{-i}(k); c(k)) - \lambda,\]
where
\[(pb_{1},..,pb_{n})\,\epsilon\, D^{F} \mbox{ and } pb_{i}(k) \ge \lambda\]

Let $\Phi^{PF}$ be the set of per-feature extensions of $\Phi$. The following properties hold for any $\pi\,\epsilon\,\Phi^{PF}$ in the light of the properties of family $\Phi$ in \cite{jm}.
\[anonymity: \pi(pb_{i};,pb_{-i});C) \mbox{ is a symmetric function of } pb_{-i}\]
\[budget-balance: \sum_{i = 1}^{n}\pi(pb_{i}; pb_{-i}; C) = \sum_{k=1}^{F} c(k)\]
\[core bounds: 0 \leq \pi(pb_{i};pb_{-i};C) \leq \sum_{k = 1}^{F} pb_{i}(k)\]
The above mentioned properties follows directly from the corresponding properties of all $\varphi\,\epsilon\,\Phi$. In this paper, we adopt a \emph{per-feature proportional cost sharing rule} that obeys the properties of anonymity, budget-balance, and core-bounds. We will explain the rule amidst our auction-based mechanism in Section 4.

In order to construct our auction-based mechanism, we define an auxiliary function $\mu_{k}$ over $R_{+}^{n}$ in lines with the work in \cite{jm}. We define $\mu_{k}$ by the following relations.
\[F1 = G1 \mbox{ if } pb_{N/i}(k) \leq jb_{i}(k);\]
\[\mu_{k}(jb_{i}(k); pb_{-i}(k)) = 0 \mbox{ if } pb_{N/i}(k) > jb_{i}(k),\]
where
\[F1 = \mu_{k}(jb_{i}(k); pb_{-i}(k);C(k)),\]
\[G1 = \varphi_{k}(jb_{i}(k) - pb_{N/i}(k);pb_{-i}(k);C(k))\]

In view of the monotonicity properties of $\varphi_{k}$, we get the following three properties of $\mu_{k}$ for the multi-feature case $\forall k = 1,...,F$.
\[\mu_{k} \mbox{ is non-decreasing in } jb_{i}(k) \mbox{ if } pb_{N/i} > 0;\]
\[A1 \geq B1 \geq C1, \forall \lambda > 0; jb_{i}(k) > 0;\]
and
\[D1 \leq E1 \mbox{ if } pb_{N/i} < jb_{i}(k),\]
where
\[A1 = \mu_{k}(jb_{i}(k); pb_{-i}(k)),\]
\[B1 = \mu_{k}(jb_{i}(k); pb_{j \ne i}(k) + \lambda,.....,pb_{n}(k)),\]
\[C1 = \mu_{k}(jb_{i}(k); pb_{-i}(k)) - \lambda,\]
\[D1 = \mu_{k}(jb_{i}(k) + \lambda; pb_{j \ne i} + \lambda,....,pb_{n}(k)),\]
\[E1 = \mu_{k}(jb_{i}(k); pb_{-i}(k)),\]
\[jb_{N}(k) = \sum_{i=1}^{n} jb_{i}(k); jb_{N/j}(k) = \sum_{i \ne j} jb_{i}(k),\]

\section{Auction Mechanism}
We devise a two stage \emph{universal} mechanism\footnote{The mechanism uses no statistical information about the distribution of other user characteristics} for implementing router deployment for public welfare. Our mechanism is based on ideas by Jackson and Moulin(1992). We aim for \emph{individual rationality, budget balance, and ex-post efficiency} without adopting the Bayesian approach to mechanism design. The authors in \cite{jm} do not consider Bayesian incentive compatibility due to some \emph{impossibility theorems} in mechanism design. However, they do mention the fact that accounting for Bayesian information about other user preferences is a rational thing to do. We can afford to use a non-Bayesian method in the present day due to the presence of social websites like Facebook, via which users can get information about each others preferences much more easily and practically without \emph{assuming} beliefs about preferences. For more details on the properties of individual rationality, budget-balance, and ex-post efficiency, the reader is referred to \cite{ngnp}\cite{grng}. Our auction mechanism is as follows.

\textbf{Stage 1:} There are $n$ agents in a locality who are interested in installing routers. Each agent $i$ simultaneously submit \emph{bids}, $B_{i}$, for the joint benefit they estimate for incorporating a certain number of router features. $B_{i}$ is a vector of the form $(jb_{i}(1),.....,jb_{i}(G_{i}))$, where $jb_{i}(k)$ denotes the joint benefit perceived by user $i$ for incorporating the $k$th feature in the router, and $G_{i}$ is the optimal number of features to be manufactured, as proposed by user $i$. The users could estimate the joint benefits via messaging in Facebook. We assume that $jb_{i}(k)$ is non-decreasing in $k$ \emph{(each feature could be in order of decreasing importance to general network users)}, for each user in the residential locality.

If $jb_{i}(1) \le c(1)$ for all $i$, the router is not built and the mechanism terminates. Otherwise, let the winner of the $k$th feature bid $\overline{jb}(k)$, and let $\overline{G}$ be the optimal number of features to be incorporated indicated by all the winning bids. Proceed to Stage 2 of the mechanism.

\textbf{Stage 2:} Users simultaneously submit individual benefit bids $ib_{i}(k)$ for each router feature $k$. The defining factors at this stage of the mechanism are the number of features each user adopts, and the cost it has to pay for its accrued benefits. There are two types of users in the residential locality: 1) those who do not win any bids in stage one of the mechanism, and 2) users who win atleast one bid. Any non-winner $i$'s adopted number of features $k^{NWU}$ is given as
\[k^{NWU} = max\{G|\alpha \ge \beta , \forall K\,\epsilon \, [1, \overline{G}]\},\]
where
\[\alpha = \sum_{k = 1}^{k = G}[ib_{N}(k) - \overline{jb}(k)],\]
\[\beta = \sum_{k = 1}^{k = K}[ib_{N}(k) - \overline{jb}(k)],\]

$ib_{N} = \sum_{j = 1}^{j = n}ib_{j}$ is the sum of individual benefit levels. We term users who win atleast one bid as `winners'. We denote the adopted number of features by a winner $i$ as $k^{WU_{i}}$, where $k^{WU_{i}}$ lies within the interval $[k^{WU_{i}}, k^{WU**_{i}}]$, if $k^{WU_{i}}\,\epsilon \,\overline{[k^{WU*_{i}}, k^{WU**_{i}}]}$, and within the interval $[\underline{k}^{WU_{i}}, k^{WU*_{i}}]$ $\bigcup \{k^{NWU}\},$ if $k^{NWU}\,\epsilon\,[k^{WU*_{i}}, k^{WU**_{i}}]$, where we have the following definition of the parameters.
\[\underline{k}^{WU_{i}} = max\{0, max\{k|ib_{N}(k) > \overline{jb}(k); I_{i}(k) = 1\}\}\]
\[k^{WU*_{i}} = max\{0, max\{k|ib_{N}(k) \ge \overline{jb}(k);I_{i}(k) = 1\}\}\]
\[k^{WU**_{i}} = min\{G, min\{k - 1| k > k^{WU*_{i}}; I_{i}(k) = 1\}\}\]
$I_{i}(k)$ is an index function which takes a value 1 if user $i$ wins the $kth$ unit and 0 otherwise.

An important part of our mechanism is cost allocation. We now describe the cost allocation mechanism for the users in the residential locality. This is done by a social network planner, who could be one amongst the locality, or be a third party who works in consultation with the SCNP to properly allocate costs according to the benefits accrued by the users.

A non-winner $i$ pays
\[\gamma - \delta,\]
where
\[\gamma = \sum_{k = 1}^{k = k^{NWU}} \mu_{k}(\overline{ib}(k), ib_{-i}(k)),\]
\[\delta = \sum_{k = k^{NWU} + 1}^{k = \overline{G}}[\overline{ib}(k) - ib_{N/i}(k) - \mu_{k}(\overline{ib}(k), ib_{-i}(k))]\]

The first part of the payment expression denotes amount of features adopted by the winner, whereas the second part is the amount of compensation received from other users in the network. The compensation could be in the form of deductions to the monthly bill that would have arrived without the compensations. The compensations resembles the price of overbidding in the first stage of the game mechanism.

A winner who atleast wins one bid in Stage 1 pays
\[\alpha1 + \alpha2 - \alpha3 + \alpha4,\]
where
\[\alpha1 = \sum_{k\epsilon L_{i}^{S}}[c(k) - \sum_{j\epsilon N/i}\mu_{k}(\overline{ib}(k), ib_{-j}(k))],\]
\[\alpha2 = \sum_{k \not \in L_{i}^{S}; \,k \le k^{WU_{i}}}\mu_{k}(\overline{ib}(k), ib_{-i}(k)),\]
\[\alpha3 = \sum_{k\epsilon\,(k^{WU_{i}}, \overline{G}); \,k \not \in L_{i}^{F}}[\overline{ib}(k) - ib_{N/i} - \mu_{k}(\overline{ib}(k), ib_{-i}(k))],\]
\[\alpha4 = \sum_{k\epsilon L_{i}^{F}}[\sum_{j\epsilon N/i}[\overline{ib}(k) - ib_{N/j}(k) - \mu_{k}(\overline{ib}(k), ib_{-j}(k))]],\]
where we have $L_{i}^{S}, L_{i}^{F}$ defined as
\[L_{i}^{S} = \{k|k \le k^{WU_{i}}, I_{i}(k) = 1; ib_{N} \ge \overline{ib}(k)\}\]
\[L_{i}^{F} = \{k|I_{i}(k) = 1; \,k \not \in L_{i}^{S}\}\]
We note that the first part of the pay of user $i$ is its cost-share as a winner and the second part is its cost share as a non-winner. The third part is the compensation received as a non-winner and the fourth part is the fine that the user pays. The $\mu$ function is a family of auxiliary functions that we define over $R^{n+}$ which define the cost allocation rules of the social planner and satisfies budget balance and individual rationality.

We summarize the result of this paper via the following theorem. \\
\textbf{Theorem.} \emph{At every sub-game perfect Nash equilibrium (SPNE) in undominated strategies, 1) the router is supplied by the SCNP with the optimal number of features, 2) the costs amongst the users are distributed according to a proportional cost sharing mechanism which is successful, budget-balanced, ex-post efficient, and individually rational, 3) the highest first stage bids equals the correct joint benefit, and 4) the second stage bids reveal a user's true valuation for each feature.}

\emph{Proof.} The proof of the above theorem follows from the following four lemmas, the proofs of which are in the Appendix. \\
\textbf{Lemma 1.} \emph{A non-winner has a unique dominant strategy to report its individual valuations for each feature truthfully.}\\
\textbf{Lemma 2.} \emph{A winner has a dominant strategy to report truthfully on its non-winning features.} \\
\textbf{Lemma 3.} \emph{Given that no user uses a weakly dominated strategy and that Lemmas 1 and 2 hold strictly, the best bidding strategy for any winner} $i$ \emph{is: 1)} $\forall k$ \emph{such that} $I_{i}(k) = 0$, $ib_{i}^{optimal}(k) = pb_{i}^{k}$\emph{and 2)}$\forall k$ \emph{such that} $I_{i}(k) = 1$, $ib_{i}^{optimal}(k) = max\{0, \overline{jb}(k) - pb_{N/j}(k)\}$.\\
\textbf{Lemma 4.} \emph{In stage 1 of our auction mechanism game, user} $i's$ \emph{optimal bid for feature} $k$ \emph{is given by} $ib_{i}^{optimal}(k)$, \emph{which lies in the interval} $[0, \overline{ib}_{-i}(k))$ \emph{if} $\overline{ib}_{-i}(k) \ge pb_{N}(k)$ \emph{and} $k \le F$, \emph{and is equal to} $max\{c(k), \overline{ib}_{-i}(k)\}\, + \,\epsilon$ \emph{if} $\overline{ib}_{-i}(k) \ge pb_{N}(k)$. \emph{Here} $\epsilon$ \emph{is the smallest increment allowed and} $F = max\{k|pb_{N}(k) > c(k)$. \emph{A user does not bid for the} $kth$ \emph{feature if} $k > F$. \
\section{Conclusion}
In this paper, we have developed an universal, two-stage, auction-based mechanism to catalyze efficient deployment of multi-feature routers in a wireless social community network. Our mechanism is simple-to-implement, successful, budget-balanced, ex-post efficient and individually rational. Our proposed mechanism accounts for heterogeneous user preferences towards different router features and comes up with the optimal \emph{(feature-set, user costs)} router blueprint that satisfies each user in a locality. Our work takes an important step towards motivating community users in buying WSCN routers and in turn contributing to the increase in mobile and ubiquitous wireless WiFi coverage and QoS improvement.

\section{Appendix}
In this section, we prove the theorem stated in Section IV. Our proof is divided into four parts in the form of lemmas. The statement of the theorem follows from the four lemmas. \\

\textbf{Lemma 1.} \emph{A non-winner has a unique dominant strategy to report its individual valuations for each feature truthfully.}\\

\emph{Proof.} A non-winner $j$ affects its payoff via $k^{NWU}$. Given $ib_{-j}$ of other agents and the winning bids $(\overline{jb}(1),....,\overline{jb}(k))$ in stage 1 of the auction mechanism, $k^{NWU}$ is a function of $\overline{ib}_{j} = (ib_{j}(1), ......ib_{j}(k)$. We get three scenarios to deal with: 1. $k^{NWU}(\overline{ib}_{j}) = k^{NWU}(\overline{pb}_{j})$, where $\overline{pb}_{j} = (pb_{j}(1), ......pb_{j}(k)$, 2. $k^{NWU}(\overline{ib}_{j}) < k^{NWU}(\overline{pb}_{j})$, and 3. $k^{NWU}(\overline{ib}_{j}) > k^{NWU}(\overline{pb}_{j})$. In scenario 1, a user is indifferent between reporting $\overline{ib}_{j}$ and $\overline{pb}_{j}$. In scenario 2, by reporting $\overline{ib}_{j}$, user $j$ gets an utility of $U_{j}(\overline{ib}_{j}) = P1 - P2 + P3$, where $P1 = \sum_{k = 1}^{k = k^{NWU}(\overline{ib}_{j})}pb_{j}(k)$, $P2 = \sum_{k = 1}^{k = k^{NWU}(\overline{ib}_{j})}\mu_{k}(\overline{ib}_{j}(k), ib_{-j}(k))$, and $P3$ is equal to $\sum_{k = k^{NWU}(\overline{ib}_{j}) + 1}^{k = \overline{G}}[\overline{ib}(k) - ib_{N/j}(k) - \mu_{k}(\overline{ib}(k), ib_{-j}(k))]$. By reporting $\overline{pb}_{j}$, user $j$ gets an utility of $U_{j}(\overline{pb}_{j}) = P1' - P2' + P3'$, where $P1' = \sum_{k = 1}^{k = k^{NWU}(\overline{pb}_{j})}pb_{j}(k)$, $P2'$ is equal to $\sum_{k = 1}^{k = k^{NWU}(\overline{pb}_{j})}\mu_{k}(\overline{ib}_{j}(k), ib_{-j}(k))$, and $P3'$ is equal to $\sum_{k = k^{NWU}(\overline{pb}_{j}) + 1}^{k = \overline{G}}[\overline{ib}(k) - ib_{N/j}(k) - \mu_{k}(\overline{ib}(k), ib_{-j}(k))]$. We know that $max\{G|\sum_{k = 1}^{k = G}[pb_{j}(k) + ib_{N/j}(k) - \overline{jb}(k)] \ge \sum_{k = 1}^{k = K}[pb_{j}(k) + ib_{N/j}(k) - \overline{jb}(k)],\forall K\,\epsilon \, [1, \overline{G}]\}$.Thus, $U_{j}(\overline{ib}_{j}) - U_{j}(\overline{pb}_{j}) \ge 0$. Therefore, in scenario 2, an user prefers a truthful report $\overline{pb}_{j}$ to a report $\overline{ib}_{j}$. In case of scenario 3, $U_{j}(\overline{ib}_{j}) - U_{j}(\overline{pb}_{j}) = \sum_{k = k^{NWU}(\overline{pb}_{j}) + 1}^{k = k^{NWU}(\overline{pb}_{j})}[pb_{j}(k) + ib_{N/j}(k) - \overline{jb}(k)]$. We again know that $max\{G|\sum_{k = 1}^{k = G}[pb_{j}(k) + ib_{N/j}(k) - \overline{jb}(k)] \ge \sum_{k = 1}^{k = K}[pb_{j}(k) + ib_{N/j}(k) - \overline{jb}(k)],\forall K\,\epsilon \, [1, \overline{G}]\}$.Thus, $U_{j}(\overline{ib}_{j}) - U_{j}(\overline{pb}_{j}) \le 0$.Therefore, in scenario 3, an user prefers a truthful report $\overline{pb}_{j}$ to a report $\overline{ib}_{j}$. $\blacksquare$ \\

\textbf{Lemma 2.} \emph{A winner has a dominant strategy to report truthfully on its non-winning features.} \\

\emph{Proof.}  We divide user $i$'s report into two parts: user $i$'s report on its winning units and non-winning units represented as a pair $(\overline{ib}_{i}^{WU}, \overline{ib}_{i}^{NWU})$. Let $(\overline{pb}_{i}^{WU}, \overline{pb}_{i}^{NWU})$ denote a user's true benefit from its winning and non-winning units. We claim that any report of the form $(\overline{ib}_{i}^{WU}, \overline{ib}_{i}^{NWU})$ such that $\overline{ib}_{i}^{NWU} \ne \overline{pb}_{i}^{NWU}$ is weakly dominated by the strategy $(\overline{ib}_{i}^{WU}, \overline{pb}_{i}^{NWU})$. Given $k^{WU_{i}}$, user $i$'s cost share is independent on its non-winning features $(\overline{ib}_{i}^{NWU})$. $(\overline{ib}_{i}^{NWU})$ can only influence $k^{WU_{i}}$ through $k^{NWU}$. There are three cases to deal with: 1. when $k^{NWU}(\overline{pb}_{i}^{NWU}) \not \in (k^{WU*_{i}}, k^{WU**_{i}})$, user $i's$ choice interval will not improve by manipulating its report on its non-winning units $\overline{ib}_{i}^{NWU}$. When $k^{NWU}(\overline{pb}_{i}^{NWU}) \in (k^{WU*_{i}}, k^{WU**_{i}}]$, and $k^{NWU}(\overline{ib}_{i}^{NWU}) \not \in (k^{WU*_{i}}, k^{WU**_{i}}]$, we can easily prove by contradiction that user $i's$ optimal choice of $K^{WU_{i}}$ is not in $(k^{WU*_{i}}, k^{NWU}(\overline{pb}_{i}^{NWU}))\bigcup (k^{NWU}(\overline{pb}_{i}^{NWU}),k^{WU**_{i}}]$. In the third case, when $k^{NWU}(\overline{pb}_{i}^{NWU}) \in (k^{WU*_{i}}, k^{WU**_{i}}]$, and $k^{NWU}(\overline{ib}_{i}^{NWU}) \in (k^{WU*_{i}}, k^{WU**_{i}}]$, it can be easily shown on the lines of case 2 that $U_{i}(k^{NWU}(\overline{pb}_{i}^{NWU})) \ge U_{i}(k^{NWU}(\overline{ib}_{i}^{NWU}))$. Thus, a report $(\overline{ib}_{i}^{WU}, \overline{pb}_{i}^{NWU})$ \emph{weakly dominates} \cite{ft} any other report $(\overline{ib}_{i}^{WU}, \overline{ib}_{i}^{NWU})$ such that $\overline{pb}_{i}^{NWU} \ne \overline{ib}_{i}^{NWU})$. $\blacksquare$\\

\textbf{Lemma 3.} \emph{Given that no user uses a weakly dominated strategy and that Lemmas 1 and 2 hold strictly, the best bidding strategy for any winner} $i$ \emph{is: 1)} $\forall k$ \emph{such that} $I_{i}(k) = 0$, $ib_{i}^{optimal}(k) = pb_{i}^{k}$\emph{and 2)}$\forall k$ \emph{such that} $I_{i}(k) = 1$, $ib_{i}^{optimal}(k) = max\{0, \overline{jb}(k) - pb_{N/j}(k)\}$.\\

\emph{Proof.} We emphasize that needs to know only the joint benefit for each feature to evaluate its best response. The first part of the best reply is proved as a dominant strategy in Lemma 2. For any user $i's$ winning features, if $ib_{i}^{optimal}(k) > \overline{jb}(k) - pb_{N/j}(k)$, $i$ gets the feature and pays $c(k) - \sum_{j\epsilon N/i}\mu_{k}(\overline{ib}(k), ib_{-j}(k))]$ for it. We observe that user $i's$ payment is non-decreasing in $ib_{i}^{optimal}(k)$. Thus, if user $i$ would like to consume at least $k$ units, the best report on the $kth$ unit is $max\{0, \overline{jb}(k) - pb_{N/j}(k)\}$. Now if $ib_{i}^{optimal}(k) \le \overline{jb}(k) - pb_{N/j}(k)$, irrespective of the number of features user $i$ adopts, he needs to pay a fine for the $kth$ feature, which amounts to $\overline{ib}(k) - ib_{N/i} - \mu_{k}(\overline{ib}(k), ib_{-i}(k))$. This fine is non-increasing in $ib_{i}^{optimal}(k)$. Thus a user can minimize its fine by reporting $max\{0, \overline{jb}(k) - pb_{N/j}(k)\}$. Thus, irrespective of whether user $i$ adopts feature $k$, its best reply is $max\{0, \overline{jb}(k) - pb_{N/j}(k)\}$. $\blacksquare$\\

\textbf{Lemma 4.} \emph{In stage 1 of our auction mechanism game, user} $i's$ \emph{optimal bid for feature} $k$ \emph{is given by} $ib_{i}^{optimal}(k)$, \emph{which lies in the interval} $[0, \overline{ib}_{-i}(k))$ \emph{if} $\overline{ib}_{-i}(k) \ge pb_{N}(k)$ \emph{and} $k \le F$, \emph{and is equal to} $max\{c(k), \overline{ib}_{-i}(k)\}\, + \,\epsilon$ \emph{if} $\overline{ib}_{-i}(k) \ge pb_{N}(k)$. \emph{Here} $epsilon$ \emph{is the smallest increment allowed and} $F = max\{k|pb_{N}(k) > c(k)$. \emph{A user does not bid for the} $kth$ \emph{feature if} $k > F$. \\

\emph{Proof.} The proof follows easily from the analysis of stage 1 in \cite{jm}.\\

The proof of the theorem follows from the above lemmas.




\bibliographystyle{abbrv}
\bibliography{alluvion1}
\balancecolumns


\end{document}